\documentstyle[aps,prl,twocolumn,epsfig]{revtex}
\begin{document}
\title{\bf Adaptive learning by extremal dynamics and negative
feedback}   
 
\author{Per Bak*{$\ddagger$}** and Dante R.
Chialvo*{$\ddagger$}{$\dagger$}}
\address{{$\ddagger$}Niels Bohr Institute, Blegdamsvej 17, Copenhagen,
Denmark.
{$\dagger$}Center for Studies in Physics and Biology, The Rockefeller
University, Box 212, 1230 York Avenue, New York, NY 10021, USA. {*}Santa
Fe
Institute,
1399 Hyde Park Rd., Santa Fe, NM 87501, USA.
{**}Imperial College, 180 Queens Gate, London SW7 2BZ, UK.\\}
\date{\today }
\maketitle

\begin{abstract}
We describe a mechanism for biological learning and adaptation based on
two 
simple
principles: (I) Neuronal activity propagates only through the network's 
strongest synaptic connections (extremal dynamics), and 
(II) The strengths of active synapses are reduced if mistakes are made, 
otherwise no changes occur (negative feedback). The balancing of those
two
tendencies typically shapes a synaptic landscape with configurations
which
are barely stable, and therefore highly flexible. This allows for swift 
adaptation to new situations. Recollection of past successes is achieved
by 
punishing synapses which have once participated in activity associated 
with 
successful outputs much less than neurons that have never been
successful. 
Despite its simplicity, the model can readily learn to solve complicated 
nonlinear tasks, even in the presence of noise. In particular, the
learning time
for the benchmark parity problem scales algebraically with the problem
size N, with an 
exponent $k\sim 1.4$. \end{abstract}
\section{Introduction}
In his seminal essay, ``The Science of the Artificial''\cite{Simon} the
economist Herbert Simon suggested that biological systems, including
those
involving humans, are ``satisficing'' rather than optimising. The
process of
adaptation stops as soon as the result is deemed good enough,
irrespective of
the possibility that a better solution might be achieved by further
search. 
In reality, there is no way to find global optima in complex
environments, 
so there is no alternative to accepting less than perfect solutions 
that happen to be within reach, as  Ashby sustained in his ``Design for
a 
brain''\cite{ashby}. We shall present results on a schematic ``brain''
model 
of self-organized learning and adaptation that operates using the
principle 
of satisficing. 
The individual parts of the system,
called synaptic connections, are modified by a negative feedback process
until the output is deemed satisfactory; then the process stops. 
There is no further reward to the system once an adequate result has
been 
achieved: this is learning by a stick, not a carrot! The process starts
up 
again as soon as the situation is deemed unsatisfactory, which could
happen, 
for instance, when the external conditions change.  The negative signal
may represent hunger, physical pain, itching, sex-drive, or some other
unsatisfied 
physiological demand. 

Formally, our sceme is a reinforcement-learning algorithm (or rather
de-inforcement learning since there is no positive feedback), \cite{Barto}
where the 
strengths of the elements are updated on the bases of the signal from an
external 
critic, with the added twist that the elements (neuronal connections) do
not respond
to positive signals. 

Superficially, one might think that punishing unsuccessful neurons is
the
mirror equivalent to the usual Hebbian learning where successful
connections 
are strengthened \cite{hebb}. This is not the case. The Hebbian process, 
like any other positive feedback, continues ad infinitum, in the absence
of
some ad hoc limitation. This will render the successful synapse strong,
and all
other synapses relatively weak, whereas the negative feedback process
employed
here stops as soon as the correct response is reached. The successful
synaptic
connections are only barely stronger than unsuccessful ones. This makes
it 
easy for the system to forget, at least temporarily, its response 
and adjust to a new situation when need be. 

The synaptic landscapes are 
quite different in the two cases /cite{Araujo}. Positive reinforcement leads 
to a few
strong synapses 
in a background of weak synapses. Negative feedback leads to many
connections
of similar strength, and thus a very volatile, noncommittal structure.
Any positive feedback will limit the flexibility and hence the
adaptability
of the system. Of course, there may be instances where positive
reinforcement
takes place, in situations where hard-wired connection have to be
constructed
once and for all, without concern for later adaptation to new
situations.

The process is self-organized in the sense that no external computation
is
needed. All components in the model can be thought of as representing
known 
biological
processes, where the updating of the states of synapses takes place only
through local interactions, either with other neighboring neurons, or
with
extracellular signals transmitted simultaneously to all neurons. The
process
of suppressing synapses has actually been observed in the real brain and
is 
known as long term depression, or
LTD, but its role for actual brain function has been unclear
\cite{Barnes}. We
submit that {\em depression} of synaptic efficacy is the fundamental
dynamic 
mechanism in learning and adaptation, with LTP, the long term
potentiation of 
synapses usually associated with Hebbian learning, playing a secondary
role.

Although we did have the real brain in mind when setting up the model,
it is certainly not a realistic representation of the overwhelming
intricacies of the human brain. Its sole purpose is to demonstrate a
general
principle that is likely to be at work, and which could perhaps lead to
the
construction of better artificial learning systems. The model presented
here is
a ``paper airplane''. which indeed can fly but is completely inadequate
to 
explain the complexity of real airplanes.

Most neural network modelling so far has been concerned with the
artificial
construction of memories, in the shape of robust input-output
connections.
The strengths of those connections are usually calculated by the use of
mathematical algorithms, with no concern for the dynamical biological
processes that could possibly lead to their formation in a realistic
``in vivo'' situation. 
In the Hopfield model, memories are represented by
energy
minima in a spin-glass like model, where the couplings between Ising
spins represent synaptic strengths. If a new situation arises, the
connection
have to be recalculated from scratch. Similarly, the back-propagation
algorithm
underlying most commercial neural networks is a Newtonian optimization
process that tunes the synaptic connections to maximize the overlap
between the
outputs produced by the network and the desired outputs, based on
examples
presented to the network. All of this may be good enough when dealing
with
engineering type of problems where biological reality is of no concern,
but
we believe that this modelling gives no insight into how real brain-like
function might come about.

Intelligent brain function requires not only the ability to store
information, such as correct input output connections. It is mandatory
for the system to be able to adapt to new situations, and yet later to
recall
past experiences, in an ongoing dynamical process. The information
stored in the brain 
reflects the entire history that it has experienced, and can take
advantage of that experience. Our model illustrates explicitly how this
might take place.

The extremal dynamics allows one to define an ``active'' level,
representing
the strength of synapses connecting currently firing neurons. The
negative
response assures that synapses that have been associated with good
responses
in the past have strengths that are barely less than the active ones,
and
can readily be activated again by supprerssing the currently active
synapses.

The paper is organized as follows. The next section defines the general
problem 
in the context of our ideas. The model to be studied can be defined for
many
different geometries. In section III we review the
layered version of the model \cite{mistakes}, with a single hidden
layer. It will 
be shown how the correct connections between 
inputs are generated, and how new connections are formed when some of
the output
assignments change. In section IV we introduce selective punishment of
neurons,
such that synapses that have never been associated with correct outputs
are
punished much more severely than synapses that have once participated in
the generation 
of a good output. It will be demonstrated how this allows for speedy
recovery, and hierarchical storage, of old, good patterns.
In multi-layered networks, and in random networks, recovery of old
patterns 
takes place in terms of self-organized switches that direct the signal
to the
correct output. Also, the robustness of the circuit towards noise will
be
demonstrated. 

Section V shows that the network can easily perform more
complicated operations, such as the exclusive-OR (XOR) process, contrary
to
recent claims in the literature \cite{klemm}. It can even solve the
much more complicated parity problem in an efficient way. In the parity
problem,
the system has to decide whether the number of binary 1s among N binary
inputs is even or odd. In those problems,
the system does not have to adapt to new situations, so the success is
due to
the volatility of the active responses, allowing for efficient search of
state space without locking-in at spurious, incorrect, solutions. 
In the same section we show how the model can readily learn multi-step
tasks, 
adapt to new
multi-step tasks, and store old ones for later use, exactly as for the
simple single step problems.
Finally  section VI contains a few succinct remarks about the  most 
relevant points of this work.  
The simple programs that we have constructed can be 
down-loaded from our web-sites\cite{www}. For an in-depth discussion of 
the biological justification, we refer the readers to a recent article 
\cite{mistakes}.
\section{The problem}

\subsection{What is it that we wish to model?}
Schematically, we envision intelligent brain function as follows:

The brain is essentially a network of neurons connected with synapses.
Some of these neurons are connected to inputs from which they receive
information from the outside world \cite{outside}. The input neurons are
connected with other neurons. If those neurons receive a sufficiently
strong
signal, they fire, thereby affecting more neurons, and so on.
Eventually, an
output signal acting on the outside world is generated. All the neurons
that
fired in the entire process are ``tagged'' with some chemical for later
identification \cite{tagging}. The action on the outside is deemed
either good 
(satisfactory) or bad (not satisfactory) by the
organism. If the output signal is satisfactory, no further action takes
place. 

If, on the other hand, the signal is deemed unsatisfactory, a global
feedback signal - a hormone, for instance - is fed to all neurons
simultaneously. Although the signal is broadcast democratically to all
neurons, only the synapses that were previously tagged because they
connected
firing neurons react to the global signal. They will be suppressed,
whether or
not they were actually responsible for the bad result. Later, this may
lead to a
different routing of the signals, so that a different output signal may
be
generated. The process is repeated until a satisfactory outcome is
achieved,
or, alternatively, until the negative feedback mechanism is turned off,
i.e.,
the system gives up. In any case, after a while the tagging disappears.

The time-scale for tagging is not related to the time-scale of
transmitting
signals in the brain but must be related to a time scale of events in
the real outside world, such as a realistic time interval between
starting to
look for food (opening the refrigerator) and actually finding food and
eating
it. It is measured in minutes and hours rather than in milliseconds.

All of this allows the brain to discover useful responses to inputs, to
modify
swiftly the synaptic connection when the external situation changes,
since the
active synapses are usually only barely stronger than some
of the inactive ones. It is important to invoke a mechanism for low
activity
in order to selectively punish the synapses that are responsible for bad
results.

However, in order for the system to be able to
recall past successes, which may become relevant again at a later point,
it is
important to store some memory in the neurons. In accordance with our
general
philosophy, we do not envision any strengthening of successful synapses.
In
order to achieve this, we invoke the principle of selective punishment:
{\it neurons which have once been associated with successful
outputs are punished much less than  neurons that have
never been involved in good decisions.}
This yields some robustness for successful patterns with respect to
noise, and 
also helps
constructing a tool-box of neuronal patterns stored immediately below
the active level, i. e. their inputs are slightly insufficient to cause
firing.
This ``forgiveness'' also makes the system stable with respect to random
noise - a good synapse that fires inadvertently because of occasional
noise is
not severely punished. Also, the extra feature of forgiveness allows for
simple and efficient learning of sequential patterns, i. e. patterns
where
several specific consecutive steps have to be taken in order for the
output
to become successful, and thus avoid
punishment. The correct last steps of will not be forgotten when the
system
is in the process of learning early steps.

In the beginning of the life of the brain, all search must necessarily
be arbitrary, and the selective, Darwinian, non-instructional nature of
the
process is evident. Later, however, a tool-box of useful connections has
been
build up, and most of the activity is associated with previously
successful
structures - the process appears to be more and more directional, since
fewer
and fewer mistakes are committed.

Roughly speaking, the sole function of the brain is to get rid of
irritating negative feedback signals by suppressing firing neurons, in
the hope
that better results may be achieved that way. A state of inactivity,
or Nirvana, is the goal! A gloomy view of Life, indeed! The process is
Darwinian, in the sense that unsuitable synapses are killed, or at least
temporarily suppressed, until perhaps in a different situation they may
play a
more role. There is no direct ``Lamarckian'' learning by instruction,
but only
learning by negative selection.

It is important to distinguish sharply between features that must be
hardwired, i. e. genetically generated by the Darwinian evolutionary
process,
and features that have to be self-organized, i. e., generated by the
intrinsic
dynamics of the model when subjected to external
signals. Biology has to provide a set of more or less randomly connected
neurons, and a mechanism by which an output is deemed unsatisfactory, 
a ``Darwinian good selector'', transmitting a signal to all neurons
(or at least to all neurons in a
sector of the brain). It is absurd to speak of meaningful brain
processes if the purpose is not defined in advance. The brain cannot
learn to
define what is good and what is bad. In our model this is given at the
outset.
Biology also must provide the
chemical or molecular mechanisms by which the individual neurons react
to
this signal. From there on, the brain is on its own! There is no room
for further ad hoc tinkering by ``model builders''. We are allowed to
play
God, not Man!

Of course, this is not necessarily a correct, and certainly not a
complete, description of the process of self-organized intelligent
behaviour
in the brain. However, we are able to construct a specific model that
works
exactly as described above, so the scenario is at least feasible.

\subsection{So how do we actually model all of this?}
 Superficially, one would expect that
the severe limitations impose by the requirements of self-organization
will put
us in a straight-jacket and make the performance poor. Surprisingly, it
turns out that the resulting process is actually very efficient compared
with
non-self-organized processes such as back-propagation - in addition to
the fact
that it executes a dynamical adaptation and memory process not performed
by
those networks at all.

The amount of activity has to be sparse in order to solve the ``credit
(or
rather blame) assignment'' problem of identifying the neurons that were
responsible for the poor result. If the activity is high, say 50\% of
all
neurons are firing, then a significant fraction of synapses are punished
at
each time step, precluding any meaningful amount of organization and
memory. 
One could accomplish this by having a variable threshold, as in the work
by
Alstrom and Stassinopoulos\cite{Alstrom}, and by Stassinopoulos and 
Bak\cite{Stassi}. Here, we use instead ``extremal dynamics'', as was 
introduced by Bak and Sneppen (BS)\cite{BS} in a simple model of
evolution, where it resulted in a  highly adaptive self-organized
critical
state. {\it At each point in time, only a single neuron, namely the
neuron
with the largest input, fires.}

The basic idea is that at a critical state the susceptibility is
maximized,
which translates into high adaptability. In our model, the specific
state
of the brain depends on the task to be learned, so perhaps it does not
generally evolve to a strict critical state with power law avalanches
etc.
as in the BS model. Nevertheless, it always operate at a very sensitive
state
which adapts rapidly to changes in the demands imposed by the
environment.

This ``winner take all'' dynamics has support in well documented facts
in neurophysiology. The mechanism of  lateral inhibition could be the
biological mechanism implementing extremal dynamics. The neuron with the
highest input firing rate will first reach its threshold firing
potential
sending an inhibitory signal to the surrounding, competing neurons, for
instance
in the same layer, preventing them from firing.  At the same time it
sends 
an excitatory signal to other neurons downstream. In any case, there is
no need
to invoke a global search procedure, not allowed by the ground rules of
self-organization, in order to  implement the extremal dynamics. The
extremal
dynamics, in conjunction with the negative feedback, allows for
efficient
credit assignment.

One way of visualizing the process is as follows. Imagine a pile of sand
(or a river network, if you wish). Sand is added at designated input
sites,
for instance at the top row. Tilt the pile until one grain of sand
(extremal dynamics) is toppling, thereby affecting one or more
neighbors. We
then tilt the pile again until another site topples, and so on.
Eventually, a
grain is falling off the bottom row. If this is the site that was deemed
the
correct site for the given input, there are no modifications to the
pile.
However, if the output is incorrect, then a lot of sand is added along
the
path of falling grains, thereby tending to prevent repeat of the
disastrous
result. Eventually the correct
output might be reached.
If the external conditions change, so that another output is correct,
the sand, of course, will trickle down as before, but the old output is
now
deemed inappropriate. Since the path had just been successful, only a
tiny
amount of sand is added along the trail, preserving the path for
possible later
use. As the process continues, a complex landscape representing the past
experiences, and thus representing the memory of the system, will be
carved 
out.

\section{The model}
\subsection{The simplest layered model}
\begin{figure}[htbp]
\centering\psfig{figure=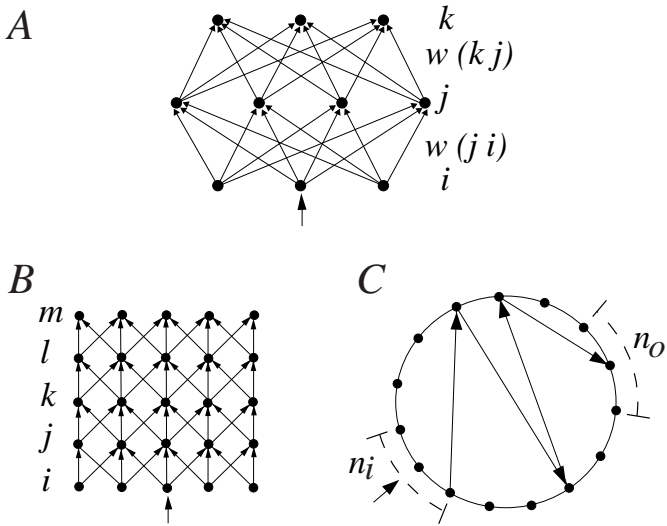,width=3.45truein}
\vspace{.25in} 
\caption{\footnotesize{Topology and notation for the three geometries of
the
model.
A) The simplest layered model with input layer $i$, connected via
synapses $w(j~i)$ to all nodes $j$ in the middle layer, which, in turn,
are
connected to all output neurons $k$ by synapses $w(k~j)$.
B)The lattice version is similar to the layered case except that
each node connects forward only with a few (three in this case)
of the neurons in the adjacent layer.
C) The random network has N neurons, $i$, each  connected with $n_c$
other
neurons $j$, with synaptic strengths $w(j~i)$ 
(only a couple are shown). Some of them, ($n_i$) are preselected as
input 
and some ($n_o$) as output neurons.
A maximum number ($t_f$) of firing is allowed in order to reach the
output.
} }
\label{fig:one}
\end{figure}

In the simplest layered version, treated in details in
Ref.\cite{mistakes}, the setup  is as follows (Fig.\ \ref{fig:one}).
There is a 
number of input cells, an intermediate layer  of ``hidden'' neurons, and
a layer
of output neurons. Each of the input neurons, $i$ is connected with each
neuron
in the middle layer, $j$, with synaptic strength  $w(j i)$ . Each hidden
neuron, in turn, is connected with each output
neuron, $k$ with synaptic strength $w(k j)$. Initially, all the
connection
strengths are chosen to be random, say with uniform distribution between
0 and 1. Each input signal consists (for the time being) of a single
input
neuron firing. For each input signal, a single output neuron represents
the
pre-assigned correct output signal, representing the state of the
external
world. The network must learn to connect each input with the proper
output for any arbitrary set of assignments, called a map. The map could
for
instance assign each input neuron $i$ to the output neuron with the same
label.
(In a realistic situation, the brain could receive a signal that there
is
some itching at some part of the body, and an output causing the fingers
to scratch at the proper place must be generated for the signal to
stop).
At each time step, we invoke ``extremal dynamics'' equivalent with a
``winner
take all'' strategy: only the neuron connected with the largest synaptic
strength to the currently firing neuron $i$ will fire at the next time
step.

The entire dynamical process goes as follows:\\

i)  An input neuron $i$ is chosen to be active.\\

ii) The neuron $j_m$ in the middle layer which is connected with the
input
neuron with the largest $w(j i)$ is
firing.\\

iii) Next, the output neuron $k_m$ with the maximum $w(k j_m)$ is
firing.\\

iv) If the output $k$ happens to be the desired one, {\bf nothing }is
done,\\

v)otherwise, that is if the output is not correct,
 $w(k_m j_m)$ and $w(j_m i)$ are both depressed by an
amount $\delta$, which could either be a fixed amount, say 1, or a
random
number between 0 and 1.\\

vi) Go to i). Another random input neuron is chosen and the process is
repeated.\\

That is all! The constant $\delta$ is the only parameter of the model,
but since only relative values of synaptic strengths are important, it
plays
no role. If one finds it un-aesthetic that the values of the connections
are
always decreasing and never increasing, one could raise the values of
all
connections such that the value of the largest output synaptic strength
for each neuron is $1$. This has no effect on the dynamics.

We imagine that the synapses $w(k_m~j_m)$ and $w(j_m~i)$ connecting all
firing neurons are ``tagged'' by the activity, identifying them for
possible subsequent punishment. In real life, the tagging
must last long enough to ensure that the result of the process is
available
- the time-scale must match typical processes of the environment rather
than
firing rates in the brain. If a negative feedback is received all the
synapses
which were involved and therefore tagged are punished, whether or not
they
were responsible for the bad result. This is democratic but, of course,
not
fair. We cannot imagine a biologically reasonable mechanism that permits
identification of synapses for selective punishment (which could of
course be 
more efficient) as is assumed in most
neural network models. The use of extremal dynamics solves the crucial
credit assignment problem, which has been a stumbling block in previous
attempts to model self organized learning, in a simple and elegant way.

Eventually, the system learns to wire each input to its correct output
counterpart. The time that it takes is roughly equal to the time that a
random 
search for each input would take. Of course, no general search process
could in
principle be faster \cite{nofreelunch} in the absence of any
pre-knowledge of the
assignment of output neurons. It is important to have a large number of
neurons 
in the middle layer in order to prevent the different paths to
interfere, and
thus destroy connections that have already been correctly learned.

\begin{figure}[htbp]
\centering\psfig{figure=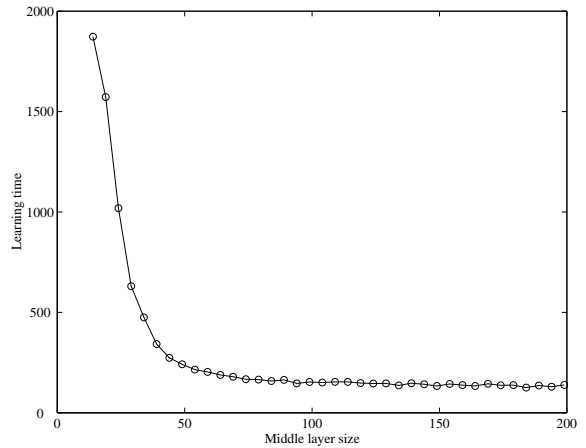,width=3.00truein}
\vspace{.25in} 
\caption{\footnotesize{The time to learn a given task decreases when the
number of neurons in the middle layer is increased. Data points are 
averages from 1024 realizations.
}}
\label{fig:morebetter} 
\end{figure}
Figure \ref{fig:morebetter} shows the results from a simulated layered
system 
with 7 input and 7 output nodes, and a variable number of intermediate
nodes. 
The task was simply to connect each input with one output node (it does
not 
matter which one). In each step we check if the seven pre-established 
input-output pairs were learnt and compute over many realizations the
average 
time to learn all input-output connections. The figure shows how the 
average learning time decreases 
with the number of hidden neurons. More is better! Biologically,
creating a 
large number of more or less identical neurons does not require more
genetic 
information than creating a few, so it is cheap. On the other hand, the
set-up
 will definitely loose in a storage-capacity beauty contest with
orthodox neural 
networks - that is the price to pay for self-organization! We are not
allowed
to engineer non-interfering paths - the system has to find them by
itself.

At this point all that we have created is a biologically motivated robot
that
can perform a random search procedure that stops as soon as the correct
result
has been found. While this may not sound like much, we believe that it
is a
solid starting point for more advanced modelling.

We now subject the model to a new input-output map. This reflects that
the external realities of the organism have changed, so that what was
good
yesterday is not good any more. Food is to be found elsewhere today, and
the
system has to adapt. Some input-output connections may still be good,
and the 
synapses connecting them are basically left alone. However, some outputs
which 
were deemed correct yesterday are deemed wrong today, so the synapses
that 
connected those will immediately be punished. A search process takes
place as 
before in order to find the new correct connections.

\begin{figure}[htbp]
\centering\psfig{figure=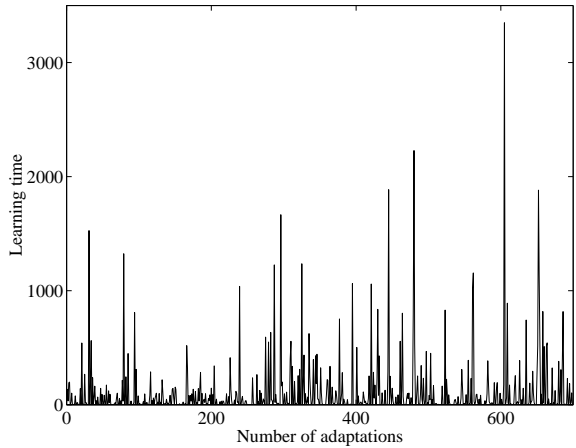,width=3.00truein}
\vspace{.25in} 
\caption{\footnotesize{Adaptation times for a sequence of 700
input-output maps.
The number of unsuccessful attempts to generate the correct input-output 
connections is shown (A random network geometry was chosen, but the
result is 
similar for the other geometries considered.)
 }}
\label{fig:wrong}
\end{figure}
Figure\ \ref{fig:wrong} shows the time sequence of the number of
``wrong''
input-output connections, i. e., which is a measure of the re-learning
time,
when the system is subjected to a sequence of different input-output 
assignments. For each re-mapping, each input neuron has a new random
output 
neuron assigned to it. In general, the re-learning time is roughly 
proportional to the number of new input-output assignments that have
changed, 
in the limit of a very large number of intermediate neurons. If the
number of 
intermediate neurons is 
small, the re-learning time will be longer because of ``path
interference'' 
between the connections. In a real world, one could imagine that the
relative 
amount of changes that would occur from day to day is small and
decreasing, so
that the re-learning time becomes progressively lower.

Suppose now that after a few new maps, we return to the original
input-output
assignment. Since the original successful synapses have been weakened, a
new
pathway has to be found from scratch. There is no memory of connections
that
were good in the past. The network can learn and adapt, but it can not
remember responses that were good in the past. In sections 4 and 6 we
shall 
introduce a simple remedy for that fundamental problem, which does not
violate
our basic philosophy 
of having no positive feedback.

\subsection{Lattice geometry}
The set-up discussed above can trivially be generalized to include more
intermediate layers. The case of multi-layers of neurons that are not
fully connected with the neurons in the next layer is depicted in figure
1B.
Each neuron in the layer connects forward to three others in the next
layer.
The network operates in a very similar way: a firing neuron in one
layer causes firing of the neuron with the largest connection to that
neuron in the subsequent layer and so on, starting with the input neuron
at
the bottom. Only when the signal reaches the top output layer will
all synapses in the firing chain be punished, by decreasing their
strength
by an amount $\delta$ as before, if need be. Interestingly,
the learning time {\em does not} increase as the number of layers
increases.
This is due to the ``extremal dynamics'' causing the speedy
formation of robust ``wires''. In contrast, the learning time for
back-propagation networks grows exponentially with the number of layers
-this
is one reason that one rarely sees backprop networks with more than one
hidden
layer.

\subsection{Random geometry}
In addition to layered networks, one can study the process in a random
network, 
which may represent an actual biological system better. Consider an
architecture
where each of $n$ neuron is arbitrarily connected to a number $n_c$ of
other 
neurons with synaptic strengths $w(j~i)$. A number of neurons
($n_i$ and $n_o$) are arbitrarily selected as input and output neurons,
respectively. Again, output neurons are arbitrarily assigned to each
input neuron. Initially, a single input neuron is firing. Using extremal
dynamics, 
the neuron that is connected with the input neuron with the largest
strength
is then firing, and so on. If after a maximum number of firing events
$t_f$
the correct output neuron has not been reached, all the synapses in
the string of firing neurons are punished as before. Summarizing, the
entire
dynamical process is as follows:\\

i) A single input neuron is chosen.\\

ii) This neuron is connected randomly with several others, and the one
which is connected with the largest synaptic strength fires. The
procedure is repeated a prescribed maximum number of times $t_f$,
thereby
creating and labelling a chain of firing neurons.\\

iii) If, during that process, the correct output has not been reached,
each synapse in the entire chain of firings is depressed an amount
$\delta$.\\

iv) If the correct output is achieved, there is no plastic modification
of the neurons that fired. Go to i)\\

A system with $n=200$, $n_i=n_o=5, n_c= 10$ behaves like the layered
structure
presented above (and is actually the one shown in the figure. This 
illustrates the striking development of an organized
network structure even in the case where all initial connections are
absolutely
uncorrelated. The model creates wires connecting the correct outputs
with
the inputs, using the intermediate neurons as stepping stones.

\section{Selective punishment and remembering old successes}

We observed that there was not much memory left the second time around,
when an old assignment map was re-employed - the task had to be
re-learned
from scratch.  This turns out to be much more than a nuisance, in
particular 
when the
task was complicated, like in the case of a random network 
with many intermediate neurons, where the search became slow.

We would like there to be some memory left from previous successful
experiences, so that the earlier efforts would not be completely wasted.

There is an analogous situation in the immune system, where the
lymphocytes can recognize an invader faster the second time around. The 
location
and activation of memory in biological systems is an important, but
largely 
unresolved problem. Speaking about the immune system, it has in fact
been 
suggested in a series of remarkable papers by Polly Matzinger that the
immune 
system is only activated in the presence of ``danger'' \cite{polly}. 
This is the
equivalent of our learning by mistakes. In fact, Matzinger realizes that
the 
identification of danger has to be pre-programmed in the innate immune
system, 
and must have evolved on a biological time scale- this is the equivalent
of our
``Darwinian good'' (or rather ``bad'', or ``danger''  selector or
indicator that decides if 
the organism is in a satisfactory state.

It turns out \cite{mistakes} that one single modification to the rules 
describedabove allows for some fundamental improvements of 
the system's ability to recognize old patterns:\\

iii a) When the output is wrong, a firing synapse that has at least once
been 
successful is punished much less than a synapse that has never been
successful.\\

For instance, the punishment of the ``good'' synapse could be of the
order of 
$10{^-2}$, compared with a depression of order unity for a ``bad''
synapse. 
The neuron has earned some 
forgiveness due to its past good performance. Biologically, we envision
that a 
neuron that does not receive a global feedback signal after firing,
relaxes 
its susceptibility to a subsequent negative feedback signal by some 
chemical mechanism. It is important to realize that the synapse "knows"
that 
it has been successful by the fact that it was not punished, so no
non-local information is invoked. Note that we have not, and will
not, include any positive Hebbian enforcement in order to implement
memory in 
the system - only reduced punishment.

We have applied this update scheme to both the layered and the random
version of the model. For the random model, we choose 200 intermediate
neurons,
plus 5 designated input neurons  and 5 output 
neurons. Each neuron was connected randomly with 10 other neurons. 
First, we 
arbitrarily assigned a correct output to each input, and ran the
algorithm 
above, until the map had been learned. After 
unsuccessful firings, punishment was applied; an amount of 0.001 to
previously 
successful neurons, and a random number between 0 and 1 for those that
had never
been successful. Then we arbitrarily changed one input-output
assignment, and 
repeated the learning scheme. A new random reassignment of a single
input-output
pair was introduced, and so on.
\begin{figure}[htbp] 
\centering\psfig{figure=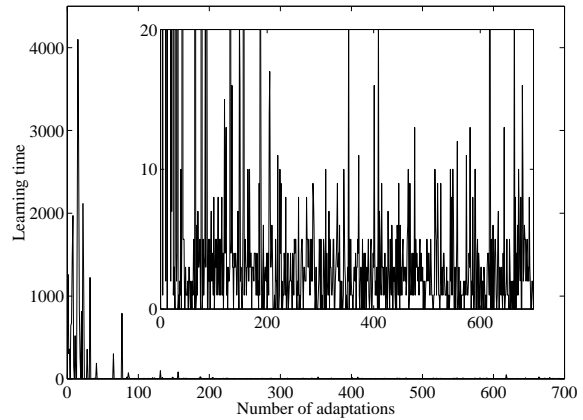,width=3.0truein} 
\vspace{.25in} 
\caption{\footnotesize{Learning time for 700 adaptations for the
random network with reduced punishment for successful synapses. Both
plots 
show the same data, but in the inset the scale magnified to better
illustrate 
the fast re-learning. The network has 5 inputs, 5 outputs, and 200
intermediate 
neurons, each connected with 10 other neurons.}} \label{fig:learntimes}
\end{figure} 
In the beginning, the learning time is large, corresponding roughly to
the time for a random search for each connection. New connections have
to be
discovered at each input-output assignment.
However, after several switches, the time for adaptation becomes much
shorter,
of the order of a few time steps. Figure \ref{fig:learntimes} shows the
time for
adaptation for hundreds of consecutive input-output reassignments. The
process 
becomes extremely fast compared with the initial learning time.
Typically,
the learning time is only 0-10 steps, compared with hundreds or
thousands
of steps in the initial learning phase.
This is because any ``new'' input-output assignment is not really new,
but has
been imposed on the system before. The entire process, in one particular
run
with 1000 adaptations, involved a total of only 38 neurons out of
200 intermediate neurons to create all possible input-output
connections,
and thus all possible maps.

In order to understand this, it is useful to introduce the concept of
the
``active level", which is simply the strength of the strongest synaptic
output connection from the neuron which has just been selected by the
extreme dynamics. For simplicity, and without changing the firing
pattern whatsoever, we can normalize this strength to unity. The
strengths of the other output synapses are thus below the active level.
Whenever
a previously successful input-output connection is deemed unsuccessful,
the 
synapses are punished slightly, according to rule iii a), only until the
point 
where a {\it single} synapse in the firing chain is suppressed slightly
below 
the active level defined by the extremal dynamics, thus barely breaking
the 
input-output connection. Thus, connections that have been good in the
past are 
located very close to the active level, and can readily be brought back
to
life again, by suppression of firing neurons at the active level if need
be.

\begin{figure}[htbp]
\centering\psfig{figure=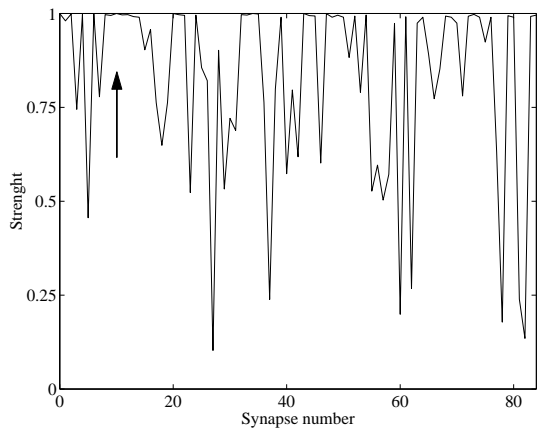,width=2.8truein}
\vspace{.25in} 
\caption{\footnotesize{Strengths of the synapses for small system with
random
connections, with 3 inputs,
3 outputs, and  20 intermediate neurons, each connected with 5 neurons.
There 
are seven active synapses with strength 1, and several synapses with
strengths 
just below the active level. Those synapses represent memories of past
successes
(such as the broken lines in Fig.\ \ref{fig:paths}.)
}}
\label{fig:landscapeA}
\end{figure}
 
Figure \ref{fig:landscapeA} shows the synaptic landscape after several  
re-learning events for a small system with 3 inputs, 3 outputs, and 20
neurons,
each connected with 5 other neurons. The arrow indicates a synapse at
the active
level, i. e., a synapse that would lead to firing if its input neuron
were 
firing. Altogether, there are 7 active synapses for that particular
simulation, representing the correct
learning of the current map. Note that there are many synaptic strengths
just 
below the active level. The memory of past successes is located in those 
synapses!

The single synapse that broke the input output connection serves as a
self-organized switch, redirecting the firing pattern from one
neuron chain to another, and consequently from one output to another.
The adaptation process takes place by employing these self-organized
switches, playing the roles of ``hub neurons'', assuring that
the correct output is reached.
\begin{figure}[htbp]
\psfig{figure=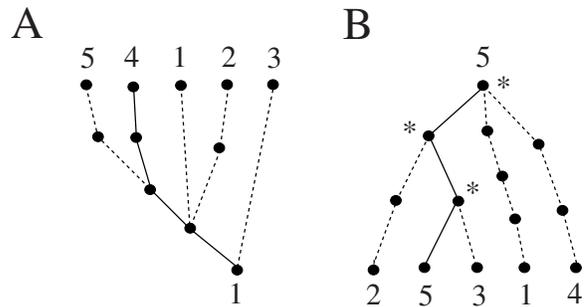,width=3.00truein}
\vspace{.25in} 
\caption{\footnotesize{a) Part of the network connecting a single output
with 
the 5 
possible inputs. The full line represents the active correct connection,
and the
broken lines represent synapses connecting with the other inputs. The
strengths 
of those synapses are barely below the active level.
b) Network connecting a single input with
all possible outputs. The synapses marked with $*$ act like
switches, connecting the input with the correct output.
}}
\label{fig:paths}
\end{figure}
 
Thus, when an input-output connection turns unsuccessful, all the
neurons
in the firing chain are suppressed slightly, and it is likely that an
old
successful connection re-appear at the active level. Perhaps that
connection 
is also unsuccessful, and will be suppressed, and another previously
successful 
connection may appear at the active level. The system sifts through old
successful connections in 
order to look for new ones.

Every now and then, there is some path interference, and re-learning
takes
longer time, indicated by the rare glitches of long adaptation times in
Figure \ref{fig:learntimes}. Also, now and then previously unused
synapses 
interfere, since the strength of the successful synapses slowly becomes
lower 
and lower. Thus even when successful patterns between all possible
input-output 
pairs have been established, the process of adaptation now and then
changes the 
paths of the connections.

Perhaps this mimics the process of thinking:

{\it ``Thinking'' is the process,
where unsuccessful neuronal patterns are suppressed by some ``hormonal''
feedback mechanism, allowing old successful patterns to emerge.  The
brain
sifts through old solutions until, perhaps, a good pattern emerges, and
the
process stops. If no successful pattern emerges, the brain panics: it
has to
search more or less randomly in order to establish a new, correct
input-output
connection.} 

The input patterns do not change during the thinking process:
one can think with closed eyes.

Figure \ref{fig:paths}a shows the entire part of the network which is
involved 
with a single input neuron, allowing it to connect with all possible
outputs.
The full line indicates synapses at the active level, connecting the
input with the correct output. The broken lines indicate synapses that
connect the input with other outputs. They are located just below the
active level. The neurons marked by an asterisk are switches, and are
responsible for redirecting the flow.

Similarly, Fig.\ \ref{fig:paths}b shows all the synapses connecting a
single 
output
with all possible inputs. The neurons with the asterisks are ``hub
neurons'', directing several inputs to a common output. Once such
neuron is firing, the output is recognized, correctly or incorrectly.
A total of only 5 intermediate neurons are involved in connecting the
output 
with all possible inputs.

Note that short-term and long-term memories are not located at, or ever
relocated to, different locations. They are represented by synapses that
are more or less suppressed relative to the currently active level
selected
by the process of extremal dynamics, and can be reactivated through
self-organized switches as described above.

The system exhibits aging at a large time scale: eventually all or most
of the
neurons will have been successful at one point or another, and the
ability to
selectively memorize good pattern disappears. The process is not
stationary.
If one does not like that, one can let the neurons die, and replaced by
fresh
neurons with random connections at a small rate. The death of neurons
causes
longer adaptation times now and then since new synaptic connections have
to
be located.

\begin{figure}[htbp]
\centering\psfig{figure=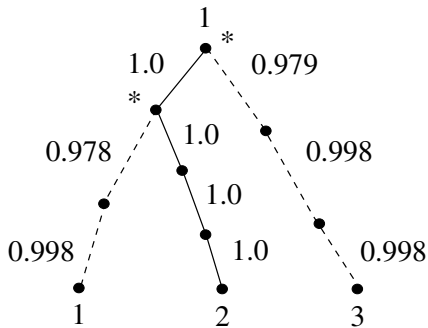,width=2.20truein}
\vspace{.25in} 
\caption{\footnotesize{Learning with noise. The diagram shows all
the synaptic connections allowing a single input neuron to connect
with all possible output neurons. The full lines show the currently
active path, and the numbers are the synaptic strengths as explained in 
text. }}
\label{fig:landscapeB}
\end{figure}

\subsection{Perfect learning with noise.}

It is also interesting to consider the effect of noise. Suppose that a
small noise $n$, randomly distributed in the interval $0 < n <
\epsilon$,
is added to the signal sent to the neurons. This may cause an occasional
wrong output signal, triggered by synapses with strengths $w(k_m j)$
that
were close to that of the correct one, i. e. the one that would be at
the active level in the absence of noise. However, those synapses will
now be
suppressed, since they lead to an incorrect result. After a while, there
will
be no incorrect synapses $w(k_m j)$ left, such that the addition of the
noise
can cause it to exceed the strength of the correct synapse $w(k_m j_m)$,
so no
further modifications will take place, and the input-output connections
will
be perfect from then on. Thus, {\it the system deals automatically with
noise!} Figure \ref{fig:landscapeB} shows all the input-output
connections
for one input neuron in a simulation with three input neurons, three
output
neurons, and a total of 50 neurons each connected with 5 neurons. The
noise level is 0.02, and the punishment of previously successful neurons
is 0.002. The numbers are the strengths of the synapses. Note that the 
incorrect synapses connected with the switches are suppressed
by a gap of at least 0.02 - the level of the noise - below the correct
ones. 
Note also 
that some of the incorrect synapses not connected with switches are much
less 
suppressed. They are cut-off by switches elsewhere and need not be
suppressed 
in order to have the signal directed to the correct output.
\begin{figure}[htbp]
\centering\psfig{figure=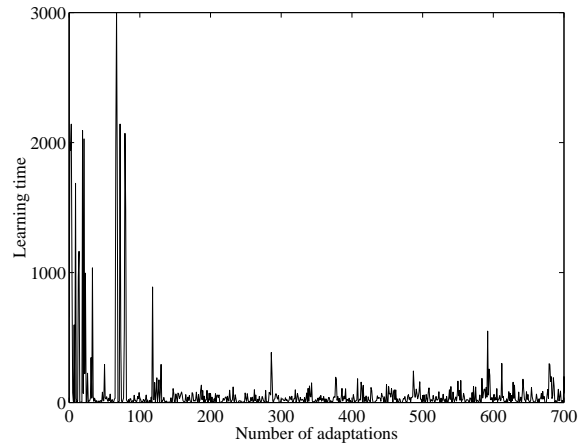,width=3.00truein}
\vspace{.25in} 
\caption{\footnotesize{ Learning times. As Fig.\ \ref{fig:learntimes},
but 
with uniform random noise of amplitude 0.02 added to the synaptic
strengths. 
Note the increase in the adaptation times. }}
\label{fig:noise}
\end{figure}
 
The price to be paid in order to have perfect learning with noise is
that
adaptation to new patterns takes longer, because the active synapses
have to be 
suppressed further to give way for new synapses. Figure \ref{fig:noise}
shows 
the learning time for 700 successive re-mappings, as in Fig.\ 
\ref{fig:learntimes}, but with noise added. Note that indeed adaptation
is much 
slower.

\section{Beyond simple wiring: XOR and sequences}
So far we have considered only simple input-output mappings where only a
single input neuron was activated. However, it is quite straightforward
to
consider more complicated patterns where several input neurons are
firing at
the same time. In the case of the layered network, we simply modify the
rule 
ii) above for the selection of the
firing neuron in the second layer as follows:\\

ii b) The neuron $j_m$ in the middle layer for which the sum of the
synaptic
connections $w(j i)$  with the active input neurons $i$ is maximum is
firing.\\

For the random network one would modify the rule for the firing of the
first 
intermediate neuron similarly.

\subsection{XOR operation}
Since the hey-days of Minsky and Papert\cite{Minsky}
who demonstrated that only linearly separable functions can be
represented by
simple -one layer- perceptrons, the ability to perform the exclusive-or
(XOR) 
operation has been considered a litmus test for the performance of any
neural 
network. How does our network measure up to the test? Following Klemm et
al. 
\cite{klemm} we choose to include three
input neurons, two of them representing the input bits for which we wish
to perform the XOR operation, and a third input neuron which is always
active. This bias assures that there is a non-zero input even when the
response
to the 00 bits is considered. The two possible outputs for the XOR
operation 
determines that the network have two output neurons.

The inputs are represented by a string of $I$ binary units $x_1, \dots,
x_{I}$,
$x_i\in\{0,1\}$. As explained in section 3,
neurons are connected by weights $w$  from each input
($j$) to each hidden ($i$) unit and from each hidden
unit to each output($k$) unit.

The dynamics of the network is defined by the following steps.
One stimulus is randomly selected out  of the four
possible (i.e.,001,101,011,111) and applied to $x_{1},x_{2},x_{3}$.
Each hidden node $j$ then receives a weighted input
$h_j=\sum_{i=1}^{I} w_{ji} x_i$.
The state is chosen according to the winner-take-all rule
i.e., the $j_m$ neuron with the largest $h_j$
fires (i. e., $x_jm=1$. Since there is only one active
intermediate neuron, the output neuron is chosen as before to be
the one connected with that neuron by the largest strength $w_{kj}$.

\begin{figure}[htbp] 
\centering\psfig{figure=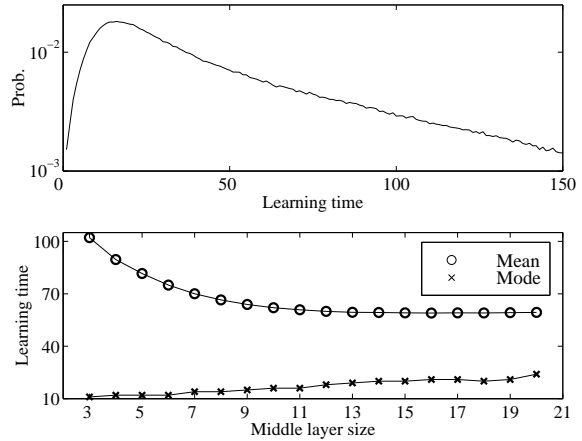,width=3.0truein,clip=true} 
\vspace{.25in}
\caption{\footnotesize{Learning the XOR problem. The top panel shows the 
distribution of learning times for a middle layer with 20 nodes. The
bottom 
panel shows the average (circles) and the mode (crosses) of the
distribution
of learning times (from $10^6$ realizations) for various sizes of the
middle 
layer.}} 
\label{fig:xorA} \end{figure}

Adaptation to changing tasks is not of interest here, so we choose to
simulate the simplest algorithm in section 3 without any selective 
punishment). As shown in Fig. \ref{fig:xorA}, networks with the minimum 
number of intermediate neurons (three for this problem) are able to
solve the 
task in as few as tens of presentations. Of course, networks with larger
middle 
layers learn significantly faster, up to an asymptotic limit which for 
this problem is reached for about 20 nodes. 
 Even in 
the present of noise, the tolerant version of the model presented above,
and in
our previous paper \cite{mistakes} allows for perfect, but slightly
slower
learning. Klemm et al introduced forgiveness in a slightly different,
and much more elaborate way, by allowing the synapses a small number
of mistakes before punishment. We do not see the advantage of this ad
hoc
scheme over our simpler original version, which also appear to be
more feasible from a biological point of view.

Indeed much 
harder problems of the same class as the XOR, can be learned by our
network 
without any modification. XOR belongs to the ``parity'' class of
problem, where 
for a string of arbitrary length $N$ there are $2^N$ realizations
composed
of all different combinations of 0 and 1's. In order to learn to solve
the 
parity problem the system must be able to selectively respond to all the 
strings with odd (or even) number of 1's (or zeros). The XOR function is
the 
simplest case with $N=2$. 

\begin{figure}[htbp]
\centering\psfig{figure=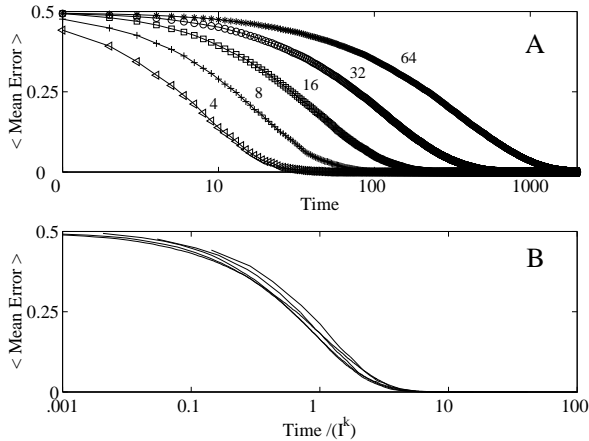,width=3.5truein}
\vspace{.25in} 
\caption{\footnotesize{Learning nonlinear problems beyond XOR. 
Curves in panel A show the time dependence of average errors for
increasingly 
harder parity functions, from order 2 (i.e., the XOR case) to
order 6. For each curve, the numbers indicate the size ($I=2{^N}$) of
the problem.
In panel B the curves shown in A are re-plotted with the time axis 
rescaled with the size of the problem, $t'= t /I^{k}$. Good data 
collapse is achieved with with $k 
\sim 1.4 $ . }} 
\label{fig:xorB}
\end{figure}

We used the same network as for the XOR problem, but now 
with increasing $N$ up to string lengths of 6. For all cases we chose a 
relatively large intermediate layer with 3000 neurons. Figure\
\ref{fig:xorB} 
shows the results of these simulations. In panel A the mean error
(calculated 
as in Klemm et al. for consistency) is the ratio between those
realizations 
which have learned the complete task and those that have not, as a
function of 
time. For each N, a total of 1024 realizations was simulated, each one
initiated
from a different random configuration of weights. Notice that the time
axis (for
presentation purposes) is in logarithmic scale. At least for the sizes
explored 
here, the network solves larger problems following not very explosive
power-law
scaling relationship. Panel B of Fig.\ \ref{fig:xorB} shows that
learning time scales 
with problem size  with an exponent $ k\sim 1.4$. In conclusion, the 
nonlinearity does not appear to introduce additional fundamental
problems into 
our scheme. 

\subsection{Generalization and feature detection}
The general focus of most neural network studies has been on the ability
of the
network to generalize, i.e., to distinguish between classes of inputs
requiring the same output. In general, the task of assigning an output
to an 
input which has not been seen before is mathematically ill-defined,
since in 
principle any arbitrary collection of inputs might be assigned to the
same 
output. Practically, one would like to map ``similar'' inputs to the
same 
output; again ``similar'' is ill-defined. We believe that similarity is
best
defined in the context of (biological) utility: similar inputs are by
definition
inputs requiring the same reaction (output) on order to be successful
(this is 
circular, of course). For a frog, things that fly requires it to stick
its 
tongue out in the direction of the flying object, so all things that fly
might 
be considered similar; there is not much need for the frog to bother
with things
that don't fly. Actually, a frog can only react to things that move as 
demonstrated in the classical paper by Lettvin, Maturana, McCulloch and
Pitts 
\cite{lettvin} almost half a century ago. Roughly, the generalization
problem 
can be reduced to the problem of identifying useful (or dangerous)
features in 
the input that have consequences for the action that should be taken.
\begin{figure}[htbp]
\centering\psfig{figure=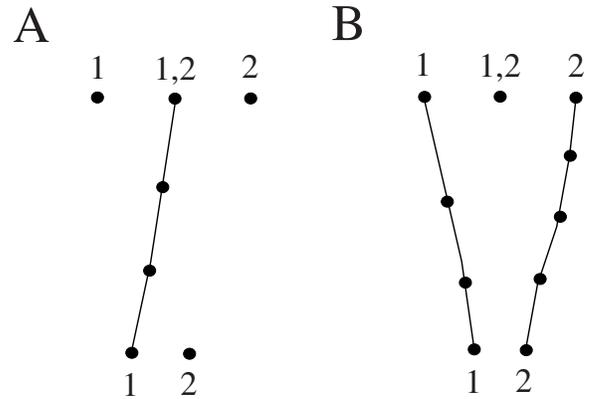,width=3.00truein}
\vspace{.25in}
\caption{\footnotesize{Two inputs, each representing two firing input
cells,
are considered. The two inputs have the input cell in the center in 
common. A) If the outputs should be the same, the
common neuron is connected with the correct output neuron. B) If the
outputs
should be different, the input neurons that are different are connected
with the two different outputs.
}}
\label{fig:generalize}
\end{figure}
So how does our network learn to identify useful features in the input?
Suppose (Fig.\ \ref{fig:generalize}), that we present two 
different inputs to, for instance, the random network, one where input
neurons 
$1$ and $2$ are firing, and another one where inputs $2$ and $3$ are
firing. 
Consider the two cases A) where the output neuron for the two inputs
should
be the same, and B) where the assigned outputs are different. 

In the case 
where the outputs should be different, say, $1$, and $2$, respectively
the 
algorithm solves the problem by connecting the input $1$ to $1$ and the
input 
$3$ to the neuron $2$ through different intermediate neurons, while
ignoring 
the input $2$. The brain identifies features in the input that are {\it 
different}. The irrelevant feature $2$ is not even ``seen'' by our
brain, 
since it have no internal representation in the form of firing
intermediate 
neurons. In the case where the assigned outputs for the two inputs are
the 
same, say $1$, the problem is solved by connecting the common input
neuron $1$
with the output neuron with a single string of synaptic connections. The 
network identifies a feature that is {\it the same} for the two inputs,
while 
ignoring the irrelevant
outputs $1$ and $3$, that are simply not registering in the brain.In a 
simulation, it was imposed that when inputs $1$ or $3$ were active
without
$2$ being active, success was achieved only if the output was not $1$:
the frog should not try to eat non-flying objects. This mechanism can 
supposedly be generalized to more complicated situations: depending on
the 
task at hand, the brain identifies useful features that allows it to 
distinguish, or not to distinguish (generalize) between inputs.

Suppose the system is subsequently presented to a pattern that in
addition
to the input neurons above includes more firing neurons. In case the
additional neurons are irrelevant for the outcome, the system will take
advantage of the connections that have already been created and ignore
the
additional inputs. If some of the new inputs are relevant, in the sense
that
a different output is required, further learning involving the new
inputs will 
take place in order to allow the system to discriminate between outputs.
We envision that this process of finer and finer discrimination between
input 
classes allows for better and better generalization of inputs requiring 
identical outputs.

The important observation to keep in mind is that the concept of
generalization 
is intimately connected with the desired function, and can not be
pre-designed. 
We feel that, for instance with respect to theories of vision, there is
an undue
emphasis on constructing general pattern detection devises that are not
based on
the nature of the specific problem at hand. Whether edges, angles,
contrasts, or
whatever are the important feature must be learned, not hardwired.

\subsection{Learning multi-step sequences.}

In general, the brain has to perform several successive tasks in order
to achieve a successful result. For instance, in a game of chess or
backgammon, the reward (or punishment) only takes place after the
completion of 
several steps. The system can not ``benefit'' from immediate punishment 
following trivial intermediate steps, no matter how much the bad
decisions 
contributed to the final poor result.

Consider for simplicity a set-up where the system has to learn to
present four successive outputs, 1, 2, 3, and 4, following a single
firing input
neuron, 1. In general, the output decision at any intermediate step will
affect 
the input at the next step. Suppose, for instance that in order to get
from one place to another in a city starting at point 1, one first has
to 
choose 
road 1 to get to point 2, and then road 2 to go to point 3, and so on.
Thus, the output represents the point reached by the action, which is
then
seen by the system and represents the following input.
We represent this by feeding the output signal to the input at the next
step. 
Thus, if output number 5 fires at an intermediate 
step, input neuron $5+1=6$ will fire at the next step: this is the outer
worlds reaction to our action.

We will facilitate the learning process by presenting the system not
only with the final problem, but also with the simpler intermediate
problems: we
randomly select an input neuron 1 to 4. If the neuron 4 is selected, the
output 
neuron 4 must respond. Otherwise the firing neurons are punished. If the
input 
neuron 3 is selected, the output neuron 3 must first fire. This creates
an input
signal at input neuron 4. Then the output 
neuron 4 must fire. For any other combination, all the synapses
participating in
the two step operation are punished. In case the input 2 is presented,
output 
neuron 2 must first fire, then output neuron 3, and finally output
neuron 4 must
fire, otherwise all synapses connecting firing neurons in the three step
process
are punished. When the input 1 is presented, the four output neurons
must
fire in the correct sequence. Of course, we never evaluate or
punish intermediate successes!

For this to work properly, it is essential to employ the selective
punishment scheme where neurons that have once participated in correct
sequences
are punished less than neurons that have never been successful, in order
for
the system to remember partially correct end games learned in the past.

In one typical run, we choose a layered geometry with 10 inputs, 10
outputs, and
20 intermediate neurons. After 4 time steps, the last step $input 4
\rightarrow 
output 4$ was
learned for the first time. After 35 time steps, the sequence $input 3
\rightarrow 
output 3 (=input4) 
\rightarrow 4$ was also learned, after 57 steps the sequence $input 2
\rightarrow 
input 3 
\rightarrow input 4 \rightarrow output 4$  was learned, and finally,
after 67 steps the 
entire sequence had been learned. These results are typical. The brain
learned 
the steps backwards, which, 
after all, is the only logical way of doing it. In chess, one has to
learn that 
capturing the king is essential before the intermediate steps acquire
any 
meaning!

In order to imitate a changing environment, we may reassign one or more
of the 
outputs to fire in the sequence. As in the previous problems, the system
will 
keep the parts that were correct, and learn the new segments. Older
sequences 
can be swiftly recalled. Finally we added uniform random noise of order  
$10^{-2}$ to the outputs; this extended the learning time in the run
above to 
193 time steps.

\section{Conclusion and Outlook}
 
The employment of the simple principles produces a self-organized,
robust and
simple, biologically plausible model of learning. It is, however,
important 
to keep in mind in which contexts these ideas do apply and in which they
do not.
The model discussed is supposed to represent a mechanism for biological
learning, 
that a  hypothetical organism could use in order to solve some of the
tasks
that must be carried out in order to secure its survival. On  the other
hand the 
model is not supposed to solve optimally any problem -  real brains are
not very 
good at that either. It seems illogical to seek to model brain function
by constructing contraptions that can perform tasks that real brains,
such as
ours, are quite poor at, such as solving the Travelling salesman
Problem. 
The mechanism that we described is not intended to be optimal, just 
sufficient for survival.

Extremal dynamics in the activity 
followed eventually by depression of only the active  synapses results
in 
preserving good synapses for a given job. In contrast to other learning
schemes, the efficiency also scales as one 
should expect from biology: bigger networks solve a given problem more
efficiently than smaller networks. 
And all of this is obtained without having to specify 
the network's structure - the same principle works well in 
randomly connected, lattices or layered networks.

In summary, the simple action of choosing the strongest and depressing
the 
inadequate synapses leads to a permanent counter-balancing which can be 
analogous to a critical state in the sense that all states in the system
are 
barely 
stable, or ``minimally'' stable using the jargon of ref. \cite{BTW}.
This 
peculiar meta-stability prevents the system from stagnating by
locking into a single (addictive) configuration from which it can be 
difficult to escape when
novel conditions arise. This feature provides for flexible 
learning and un-learning, without having to specifically include an
ad-hoc 
forgetting mechanism - it is already embedded as an integrated dynamical 
property of the system. When combined with selective punishment, the
system can
build-up a history-dependent tool box of responses that can be employed
again
in the future.

Un-learning and flexible learning are ubiquitous features of animal
learning 
as discussed recently by Wise and Murray\cite{wise}. We are not aware of 
any other simple learning scheme mastering this crucial ability.

\section{acknowledgements}
Work supported by the Danish Research Council
SNF grant 9600585.  The Santa Fe Institute receives funding from the
John D.
and Catherine T. MacArthur Foundation, the NSF (PHY9021437) and the
U.S.  Department of Energy (DE-FG03-94ER61951).


\begin{references}
\bibitem{outside}The ``outside world'' could be other parts of the
organism,
or even other parts of the brain.
\end{references}
\end{document}